\documentclass[numberedappendix,12pt,preprint,dvips]{emulateapj} 
\usepackage[]{amsmath}
\usepackage{longtable}
\usepackage{graphicx}
\usepackage{pslatex}
\newcommand{\ep}{\epsilon_{\rm p}}
\newcommand{\es}{\epsilon_{\rm s}}
\newcommand{\be}{\begin{equation}}
\newcommand{\ee}{\end{equation}}

\newcommand{\lya}{Ly$\alpha$}

\newcommand{\simgt}{\,\rlap{\lower 3.5 pt \hbox{$\mathchar \sim$}} \raise
1pt \hbox {$>$}\,}
\newcommand{\simlt}{\,\rlap{\lower 3.5 pt \hbox{$\mathchar \sim$}} \raise
1pt \hbox {$<$}\,}

\shorttitle{The increasing \lya\ optical depth at $6<z<9$}
\shortauthors{Treu et al. (2013)}

\usepackage{color}			      % using color package        - for emulateapj
\definecolor{midgray}{gray}{0.4}		% defining gray color        - for emulateapj
\definecolor{orange}{rgb}{1,0.5,0}    %        - for emulateapj

\usepackage[colorlinks=true,citecolor=midgray,linkcolor=midgray]{hyperref}        %    - for emulateapj

\newcommand{\BE}{\begin{equation}}
\newcommand{\EE}{\end{equation}}
\newcommand{\BEA}{\begin{eqnarray}}
\newcommand{\EEA}{\end{eqnarray}}

\begin{document}


\title{The changing Ly$\alpha$ optical depth in the range $6<z<9$ from MOSFIRE spectroscopy of Y-dropouts}


\author{
Tommaso Treu$^{1}$, Kasper B. Schmidt$^{1}$, Michele Trenti$^{2}$,
Larry D. Bradley$^{3}$, Massimo Stiavelli$^3$ }
\affil{$^{1}$ Department of Physics, University of California, Santa Barbara, CA, 93106-9530, USA}
\affil{$^{2}$ Institute of Astronomy and Kavli Institute for Cosmology, University of Cambridge, Madingley Road, Cambridge, CB3 0HA, United Kingdom} 
\affil{$^{3}$ Space Telescope Science Institute, 3700 San Martin Drive, Baltimore, MD, 21218, USA}

\email{tt@physics.ucsb.edu}

\begin{abstract}
We present MOSFIRE spectroscopy of 13 candidate $z\sim8$ galaxies
selected as Y-dropouts as part of the BoRG pure parallel survey.  We
detect no significant \lya\ emission (our median 1$\sigma$ rest frame
equivalent width sensitivity is in the range 2-16\AA). Using the
Bayesian framework derived in a previous paper, we perform a rigorous
analysis of a statistical subsample of non-detections for ten
Y-dropouts, including data from the literature, to study the cosmic
evolution of the \lya\ emission of Lyman Break Galaxies. We find that
\lya\ emission is suppressed at $z\sim8$ by at least a factor of three
with respect to $z\sim6$ continuing the downward trend found by
previous studies of $z$-dropouts at $z\sim7$. This finding suggests a
dramatic evolution in the conditions of the intergalactic or
circumgalactic media in just 300 Myrs, consistent with the onset of
reionization or changes in the physical conditions of the first
generations of starforming regions.
\end{abstract}

\keywords{galaxies: evolution --- galaxies: high-redshift}

% ======================================================================

\section{Introduction}
\label{sec:intro}

In the past few years our knowledge of the first galaxies has
increased stupendously. Deep imaging surveys with the Hubble Space
Telescope have pushed the frontier of Lyman Break galaxies (LBGs)
beyond redshift $z\sim10$ reaching into the epoch of cosmic
reionization \citep[e.g.,][]{Bou++10,Ell++13,Rob++13,Coe++13}. The
luminosity function of LBGs appears to evolve rapidly, with a decrease
in the number density of observed galaxies, but with faint end slopes
getting steeper
\citep[e.g.,][]{Bradley:2012p23263,Oes++13}. Similarly, narrow band
surveys on large ground based telescopes have enabled searches for
\lya\ emission yielding many candidate galaxies at comparably high 
redshift (hereafter \lya\ emitters, LAE). These studies indicate that
the amount of ionizing photons from these galaxies is sufficient to
keep the universe ionized, only if the luminosity function extends to
very faint magnitudes and the ionizing fraction is high
\citep{Trenti++10,Lor++11}.

Spectroscopic follow-up is key to further our understanding of the
physics of the first galaxies, their interactions with the surrounding
intergalactic medium, and their role in cosmic reionization.  Even
though spectroscopic follow-up of LBGs has been very successful out to
$z\sim6$, progress has been slower beyond this threshold.  Several
studies have shown that at $z\sim7$ \lya\ emission appears to be
significantly reduced with respect to $z\sim6$, consistent with a
rapid rise in the fraction of neutral hydrogen in the immediate
surroundings of these galaxies, possibly a smoking gun that we have
reached the tail-end of cosmic reionization
\citep{Kas++06,Fon++10,Pen++11,Sch++12,Ono++12,Tre++12}.

Beyond $z\sim7$, galaxies remain enshrouded in mystery, at least from
a spectroscopic point of view. Confirmation of LBGs and even of some
LAEs remain elusive \citep{Leh++10,Bun++13,Jia++13,Cap++13}. This
stems in part from technological limitations as \lya\ is redshifted
into the near infrared where traditionally spectrographs did not have
the sensitivity and multiplexing capabilities of their optical
counterparts \citep[see, e.g.,][]{Sch++12,Tre++12}.

We present here deep spectroscopic observations of a sample of 13
$z\sim8$ galaxies selected as Y-band dropouts as part of the Brightest
of Reionization Galaxies \citep[hereafter
BoRG][]{BORG1,Bradley:2012p23263}, using the new MOSFIRE
\citep{McL++08,McLean:2012p26812} spectrograph on the Keck-I
Telescope. The combination of BoRG and MOSFIRE is extremely powerful
for the study of the $z\sim8$ universe. The wide-area search of BoRG
allows us to find the brightest candidate galaxies, which also happen
to be clustered \citep{Trenti:2012p13020} in the sky and are therefore
ideal targets for the multiplexing capabilities of MOSFIRE. 

No \lya\ emission is detected down to median limiting fluxes of $0.4-0.6
\cdot 10^{-17}$ erg s$^{-1}$cm$^{-2}$ (5$\sigma$), whereas a few
detections would have been expected if the distribution of \lya\
emission had been the same as at $z\sim6$ \citep{Tre++12}. We use the
statistical framework developed by \citet{Tre++12} to perform a
rigorous analysis of the non-detections, taking into account all the
available information, and show that they imply a significant increase
in the \lya\ optical depth between $z\sim6$ and $z\sim8$.

All magnitudes are given in the AB system and a standard cosmology
with $\Omega_m=0.3$, $\Omega_{\Lambda}=0.7$ and $h=0.7$ is assumed.

%========================================================================
\section{Observations and Data Reduction}
\label{sec:borg}

%= = = = = = = = = = = = = = = = = = = = = = = = = = = = = = = = = = = = = = = = 
\begin{figure*}
\includegraphics[width=0.99\textwidth]{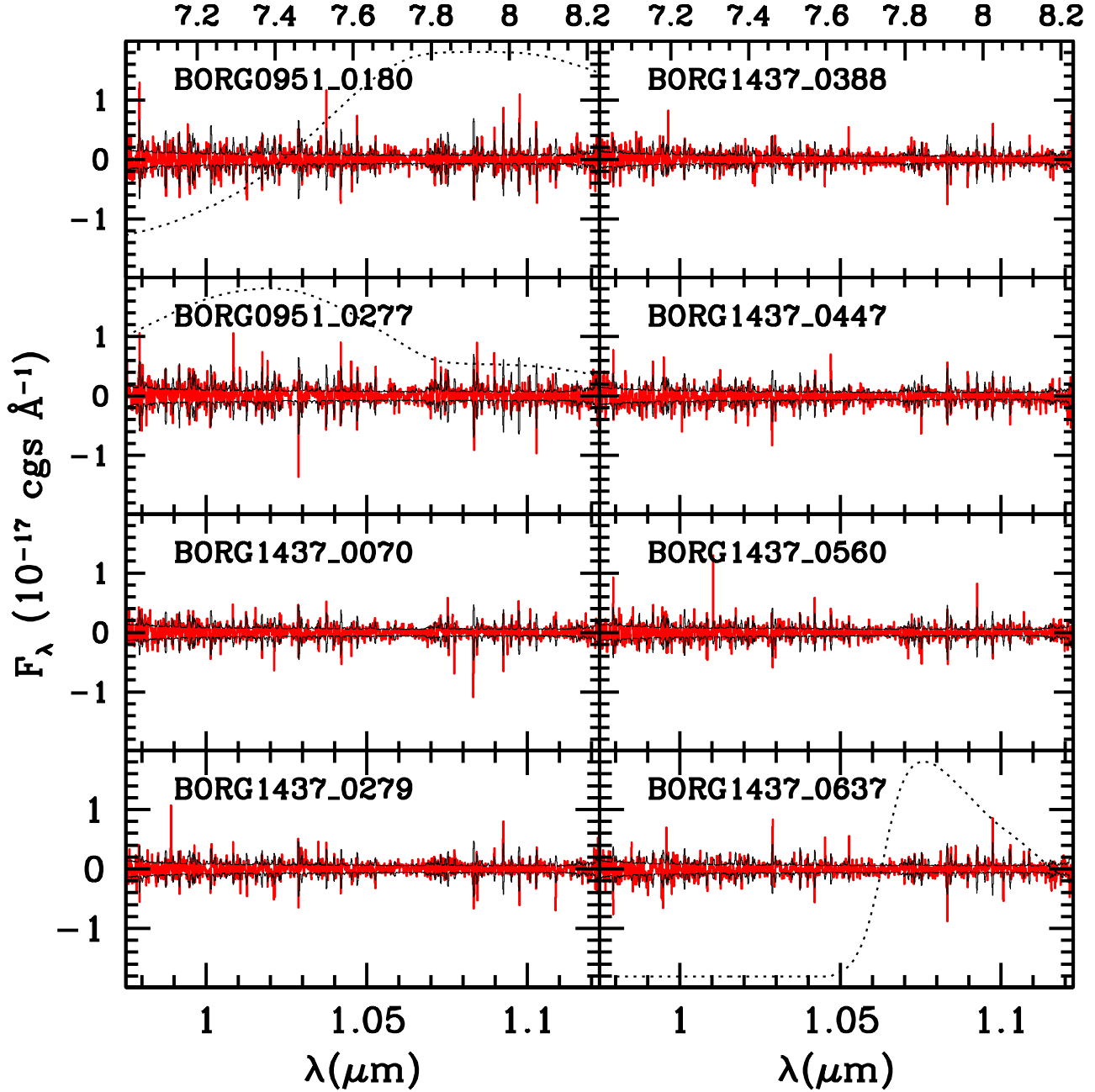}
\caption{Mosaic of the reduced MOSFIRE spectra for targets with only Y band spectroscopic data. 
The red histogram shows the actual spectrum, while the black envelope
shows the 1$\sigma$ noise level. All spectra are consistent with pure
noise. The top label shows the redshift coverage of the spectra for
\lya\ redshift. For all objects in the statistical sample (see discussion
in text) the photometric redshift posterior distribution function is
shown as the black dashed curve.  \label{fig:spec1}}
\end{figure*} 
%= = = = = = = = = = = = = = = = = = = = = = = = = = = = = = = = = = = = = = = = 

%= = = = = = = = = = = = = = = = = = = = = = = = = = = = = = = = = = = = = = = = 
\begin{figure*}
\includegraphics[width=0.99\textwidth]{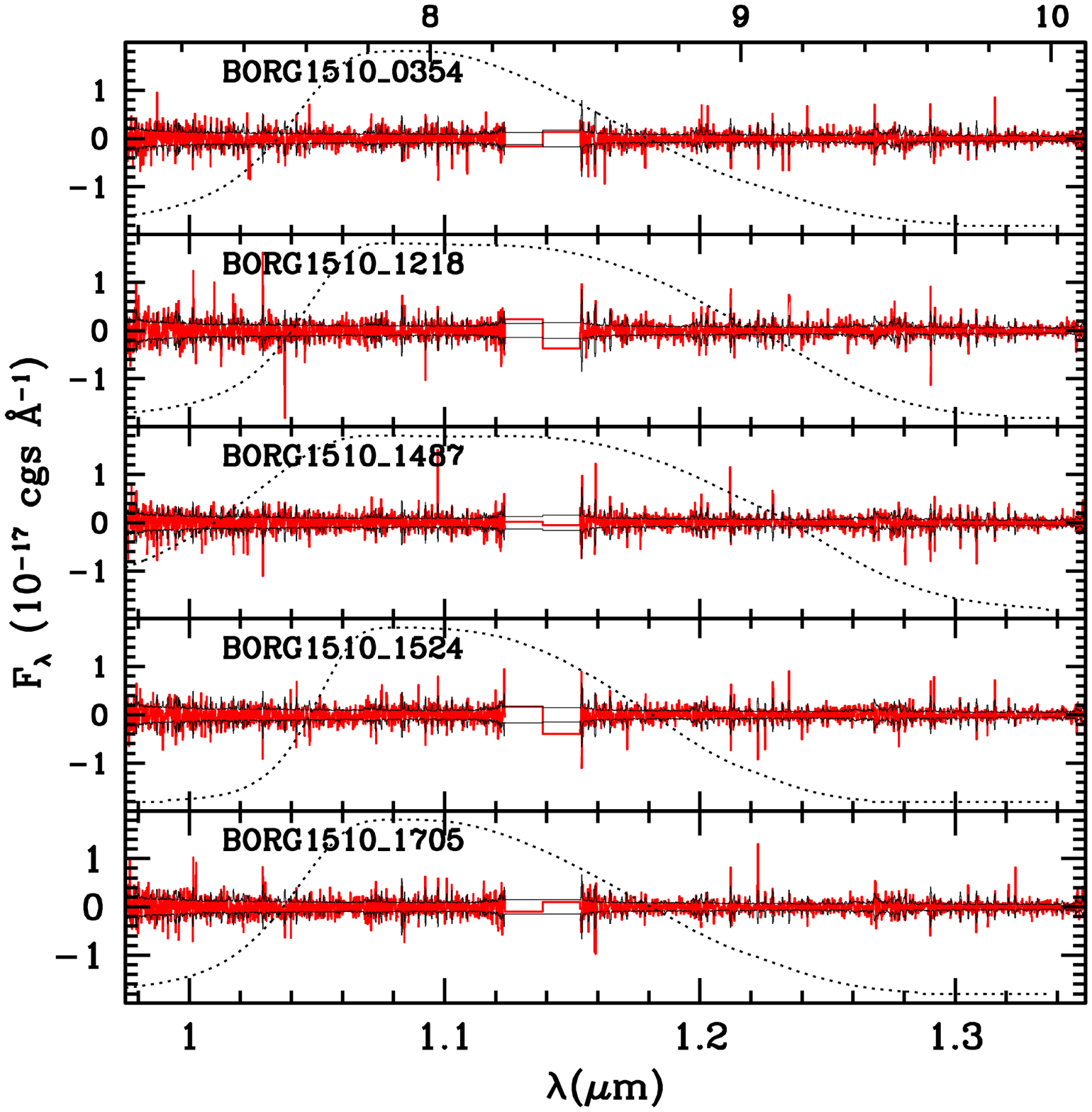}
\caption{As figure 1 for targets with both Y and J band data (the gap between the two bands is shown by the long horizontal lines).
 \label{fig:spec2}}
\end{figure*} 
%= = = = = = = = = = = = = = = = = = = = = = = = = = = = = = = = = = = = = = = = 

The spectroscopic targets were selected from BoRG as Y-band dropouts,
i.e. $z\sim8$ galaxy candidates, using \emph{Hubble Space Telescope}
(HST) data taken in the WFC3 bands F600LP/F606W, F098M, F125W, and F160W as
described by \citet{Bradley:2012p23263} and Schmidt et al. (2013; in
preparation).  The spectroscopic sample presented here consists of 13
individual targets, from 3 of the 71 BoRG fields, namely
BoRG\_0951+3304, BoRG\_1437+5043, and BoRG\_1510+1115, selected to
contain a large number of high-quality candidates. The primary
statistical sample consists of 8 dropouts detected at
high-significance ($>$5$\sigma$ in the primary detection band, plus
all the color requirements) as described by Schmidt et al. (2013; in
preparation).  Photometric redshifts for the sources in the primary
sample are shown as dashed black curves in Figures~\ref{fig:spec1}
and~\ref{fig:spec2}.  We took advantage of the multi-slit capabilities
of MOSFIRE \citep{McLean:2012p26812} to observe 5 marginal
candidates. For completeness we present results from these 5
marginal objects but do not consider them in our statistical analysis.

The near-infrared spectroscopic data presented here were obtained with
MOSFIRE on Keck-I during two nights (25-26 April 2013), in good
weather with subarcsecond seeing and clear transparency. We used slit
widths of $0\farcs7$ and a nodding amplitude of $1\farcs5$.
BoRG\_0951+3304, BoRG\_1437+5043 were observed in the Y-band, whereas
BoRG\_1510+1115 was observed both in the Y and J
bands. Table~\ref{tab:speclist} summarizes exposure times. The spatial
resolution is $0\farcs1799$ per pixel and the dispersion is 1.0855 and
1.3028 \AA/pixel in Y and J respectively. The spectral coverage is
shown in the Figures.

The data were reduced using the publicly available MOSFIRE data
reduction pipeline (DRP\footnote{http://code.google.com/p/ mosfire/}).
The output of the DRP are non-calibrated 2D spectra in units of
electrons per second per pixel.  From the 2D spectra, 1D spectra were
extracted by using a 11 pixel-wide extraction aperture centered on the
position of the target. The extracted 1D spectra were corrected for
telluric absorption using observations of the nearest V$\sim$8 A0V
Hipparcos star, for galactic extinction using the
\cite{Cardelli:1989p26704} extinction law with $R_V=3.1$, and
airmass. The absolute flux calibration was obtained from 3 bright
objects ($F098M = 20.43, 21.26, \textrm{ and } 19.57$) with known HST
magnitudes which were also observed in the MOSFIRE masks
simultaneously with the dropouts, calculating slit losses based on the
HST images.  Each night of observations was reduced separately
combining only the final spectral products at the very end. The final
calibrated spectra are shown in Figures~\ref{fig:spec1}
and~\ref{fig:spec2} and are consistent with noise. The sensitivity is
consistent with that estimated by the MOSFIRE Exposure Time
Calculator.

The non-detections allow us to rule out the possibility that the
sources in BoRG\_1510+1115 are lower redshift contaminants where the
HST continuum flux comes from emission lines like [\ion{O}{2}] and
[\ion{O}{3}] or [\ion{O}{3}] and H$\beta$
\citep{Ate++11}. However, if the continuum magnitude in F125W were due
to an emission line, it would correspond to $\sim 2-8
\cdot 10^{-17}$ erg s$^{-1}$cm$^{-2}$, detectable with our
median sensitivity of 0.4-0.6 in the same units
(5-$\sigma$). Similarly, our sensitivity is sufficient to exclude
contamination by red galaxies where a weaker line in J enhances the
break as suggested by \citet{Cap++11}. The exception would be for the
fields with only Y coverage if weak [\ion{O}{3}] fell beyond
1.1238$\mu$m ($z>1.244$) but within the F125W filter, while H$\alpha$
accounted for the flux in F160W. Longer wavelength coverage with
MOSFIRE is needed to rule out this possibility.

%= = = = = = = = = = = = = = = = = = = = = = = = = = = = = = = = = = = = = = = = 
\begin{deluxetable*}{lllccccclrr}
\tablecolumns{11} \tablewidth{0pt} \tablecaption{Summary of high redshift candidates observed with MOSFIRE}
\tablehead{\colhead{Object} & \colhead{$\alpha_\textrm{J2000}$} & \colhead{$\delta_\textrm{J2000}$} 	& \colhead{V$_{606}$} & 
\colhead{Y$_{098}$} & \colhead{J$_{125}$} & \colhead{H$_{160}$} & \colhead{Stat.} & \colhead{$t_\textrm{exp}$/hr} & 
\colhead{$\sigma_W$/\AA}}
\startdata
0951+3304\_0180			& 147.70451 & 33.06513	& $>$26.83 		& $>$26.83 	& 26.24 $\pm$ 0.27 & 26.56 $\pm$ 0.43	&  1     & 1 &  4.1 \\
0951+3304\_0277			& 147.68443 & 33.07019	& $>$26.83 		& $>$26.83 	& 25.87 $\pm$ 0.22 & 25.88 $\pm$ 0.27	&  1     & 1 &  3.0 \\
1437+5043\_r2\_0637\_T12a	& 219.21058 & 50.72601	& $>$28.10 		&  $>$28.05	& 25.76 $\pm$ 0.07 & 25.69 $\pm$ 0.08	&  1     & 3 &  2.1 \\
1510+1115\_0354			& 227.54706 & 11.23145	& $>$27.59	   	& $>$27.83 	& 27.03 $\pm$ 0.22 & 27.21 $\pm$ 0.38	&  1 & 2.1, 3  &  9.3\\
1510+1115\_1218			& 227.54266 & 11.26152	& $>$27.59	   	& $>$27.83	& 26.87 $\pm$ 0.22 & 26.64 $\pm$ 0.25	&  1 & 2.1, 3  &  8.6\\
1510+1115\_1487			& 227.53173 & 11.25254	& $>$27.59	   	& $>$27.83	& 27.60 $\pm$ 0.24 & 27.34 $\pm$ 0.28	&  1 & 2.1, 3  &  15.7\\
1510+1115\_1524			& 227.53812 & 11.25552	& $>$27.59	   	& $>$27.83 	& 26.63 $\pm$ 0.15 & 26.52 $\pm$ 0.20	&  1 & 2.1, 3  &  6.6\\
1510+1115\_1705			& 227.54008 & 11.25111	& $>$27.59	   	& $>$27.83 	& 27.00 $\pm$ 0.19 & 27.02 $\pm$ 0.28	&  1 & 2.1, 3  &  9.2\\
[0.5ex]\\\tableline\\
1437+5043\_r2\_0070\_T12e	& 219.22225 & 50.70808	& $>$28.10 		&  $>$28.05 	& 26.90 $\pm$ 0.14 & 26.94 $\pm$ 0.17	&  0     & 3 &  6.0 \\
1437+5043\_r2\_0388		& 219.23494 & 50.71960	& $>$28.10 		&  $>$28.05 	& 27.66 $\pm$ 0.24 & 27.84 $\pm$ 0.36	&  0     & 3 &  11.5 \\
1437+5043\_r2\_0560\_T12c	& 219.23092 & 50.72405	& 27.92 $\pm$ 0.31 	&  $>$28.05 	& 27.73 $\pm$ 0.23 & 27.47 $\pm$ 0.24	&  0     & 3 &  12.4 \\ 
1437+5043\_r3\_0279		& 219.18681 & 50.72723	& $>$27.94		& $>$27.82 	& 27.27 $\pm$ 0.24 & 27.46 $\pm$ 0.34	&  0     & 3 &  8.4 \\ 
1437+5043\_r3\_0447		& 219.18983 & 50.73406	& $>$27.94	   	& $>$27.82 	& 27.63 $\pm$ 0.27 & $>$27.79			&  0     & 3 &  12.3 \\
\enddata
\tablecomments{Photometry is taken from the most recent analysis by Schmidt et al. (2013; in preparation). 
and has been corrected for Galactic extinction using the
\cite{Cardelli:1989p26704} extinction law and E(B-V) of 0.01328,
0.01254, and 0.04605 for BoRG\_0951+3304, BoRG\_1437+5043 and
BoRG\_1510+1115 respectively.  The candidates in the field 1437+5043
were identified by \citet{Trenti:2012p13020} and
\citet{Bradley:2012p23263}.  The magnitude limits are 2$\sigma$ limits.
The candidates in the first part of the table (Stat=1) statisfy all
the requirements for Y-dropout selection and are the statistical
sample analyzed in this paper. The candidates below the horizontal bar
(Stat=0) were observed as slit fillers. The $t_\textrm{exp}$/hr give the total exposure time in $Y$(, $J$). 
The last column lists the median \lya\ equivalent width noise (1$\sigma$) of the MOSFIRE
spectra.  }\label{tab:speclist}
\end{deluxetable*}
%= = = = = = = = = = = = = = = = = = = = = = = = = = = = = = = = = = = = = = = =

\section{Inferences on the Ly-$\alpha$\ Optical depth}
\label{sec:inf}

\subsection{Summary of the method}
\label{ssec:method}

We apply the method introduced by \citet{Tre++12} to constrain the
distribution of equivalent width of Ly-$\alpha$ given a sample of
LBGs, exploiting all the information available. Only a brief summary
of the method is given here. The reader is referred to
\citet{Tre++12} for details and analytic expressions of the
likelihood.

As in our previous work we describe the intrinsic rest-frame
distribution in terms of the one measured at $z\sim 6$ by
\citet{SEO11} $p_6(W)$

\begin{equation}
p_6(W)=\frac{2 A}{\sqrt{2 \pi}W_{c}}e^{-\frac{1}{2}\left(\frac{W}{W_{c}}\right)^2}H(W)+(1-A)\delta(W),
\end{equation}
with W$_{c}$=47\AA, A=0.38 for sources with
$-21.75<\textrm{M}_\textrm{UV} < -20.25$ and W$_{c}$=47\AA, A=0.89 for
sources with $-20.25 < \textrm{M}_\textrm{UV}< -18.75$. A is the
fraction of emitters and $H$ is the step function. As discussed below,
1-A includes the fraction of interlopers.  Following \citet{Tre++12}
we consider two extreme cases which should bracket the range of
possible scenarios.  In the first (``patchy'') model, no \lya\ is
received from a fraction $\ep$ of the sources, while the rest is
unaffected.  The probability distribution of the equivalent width is
then given by 
\begin{equation}
\begin{split}
p_{p}(W)=\ep p_6(W)+(1-\ep)\delta(W)=\\ \frac{2A\ep}{\sqrt{2\pi}W_{c}} e^{-\frac{1}{2}\left(\frac{W}{W_{c}}\right)^2}H(W)+(1-A\ep)\delta(W).
\end{split}
\end{equation}
In the second (``smooth'') model, \lya\ is attenuated by a factor
$\es$, yielding
\begin{equation}
\begin{split}
p_{s}(W)=p_6(W/\es)/\es=\\ \frac{2A}{\sqrt{2\pi}\es W_{c}} e^{-\frac{1}{2}\left(\frac{W}{\es W_{c}}\right)^2}H(W)+(1-A)\delta(W).
\end{split}
\end{equation}

Bayes's rule gives the posterior probability of $\ep$ and $\es$
(collectively $\epsilon$) and $z$ given an observed spectrum and
continuum magnitude $m$:

\be
p(\epsilon,z_i|\{f\},m)\propto\left[\Pi_j \int dW p(f_j,m|W,z_i)p(W|\epsilon)\right] p(\epsilon)p(z_i),
\ee
where the term within square brackets is the likelihood
p(\{f\}$|\epsilon$,$z_i$,m), $f_j\in\{f\}$ are the flux measurements
in each spectral pixel and $z_i=\lambda_i/\lambda_0-1$. We adopt a uniform
prior p($\epsilon$) between zero and unity, while the prior p($z_i$)
is given by the photometric redshift.  By construction our method
takes into account the strong wavelength dependence of the
sensitivity typical of near infrared spectroscopic data. The inference
is carried out for each spectral pixel, using its noise properties
including the effects of atmospheric transmission and absorption.

The posterior on $\epsilon$ is obtained by summing over $z_i$ in
the range where p($z_i$) is non-zero. Our formalisms takes properly
into account the effects of incomplete wavelength coverage, in
deriving limits on $\epsilon$ and $z_i$. For example, if the
wavelength range only covers an interval [$z_{\rm min}$,$z_{\rm
max}$], and the prior on $\epsilon$ is uniform, then the posterior
will be:

\begin{equation}
\begin{split}
p(\epsilon|\{f\},m)\propto \sum_{z_i\in[z_{\rm min},z_{\rm max}]}p(\{f\}|\epsilon,z_i,m) p(z_i) \\ + p(\{f\}|\epsilon=0)p(z_i\notin[z_{\rm min},z_{\rm max}])
\end{split}
\end{equation} 

The normalization factor $Z$ is the Bayesian evidence and quantifies
how well the model describes the data. It can be used for selection,
by comparing evidence ratio between two models, or to decide whether
additional parameters are warranted by the data. In comparison to
other standard model selection techniques like likelihood ratio the
avdantage of the evidence is that it takes into account the entire
parameter space, thus avoiding issues of fine tuning.

The formalism described above is for an individual galaxies. For a
sample of galaxies, or for multiple independent spectra of the same
galaxies, it is sufficient to multiply the likelihoods to obtain the
total likelihood. 
 
\subsection{Results}

\begin{figure}
\includegraphics[width=0.99\columnwidth]{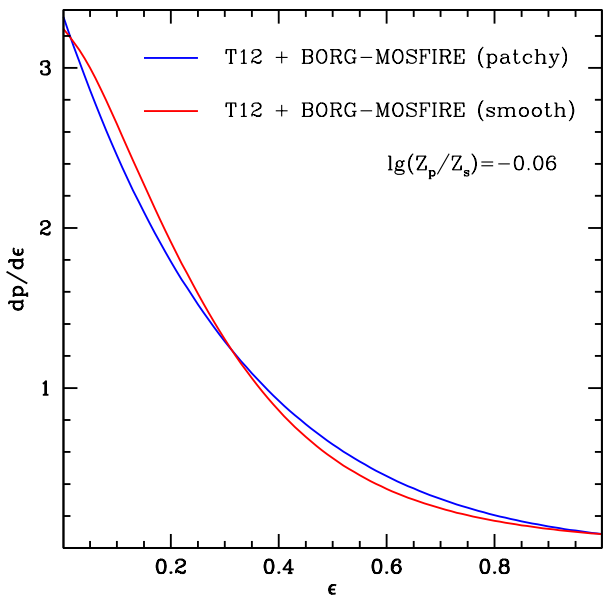}
\caption{Inference results in the context of the patchy and smooth models described in the text. The parameter $\epsilon$ describes the change of the \lya\ equivalent width distribution between $z\sim6$ and $z\sim8$. In the patchy model, at any given equivalent width, only a fraction $\ep$ of the sources that are emitting at $z\sim6$ are found to be emitters at $z\sim8$. In the smooth model the emission of each source is suppressed by a factor $\es$. The evidence ratio Z$_p$/Z$_s$ is inconclusive and does not favor any of the two models. The results shown are based on the 8 objects in the primary MOSFIRE sample presented here as well as the three spectra analyzed by \citet{Tre++12}.
 \label{fig:inference}}
\end{figure}

The key result of this paper is the posterior probability densities of
the $\ep$ and $\es$ parameters given the data, shown in
Figure~\ref{fig:inference}. This posterior probability distribution
function is based on the 8-non detections of the primary sample
presented here, plus the three non-detections presented by
\citet{Tre++12} and \citet{Sch++12} \footnote{BORG11534, A1703\_zD7 and BORG58, The latter is also part of the MOSFIRE sample presented here, so the total number of objects is 10, with one having two independent observations with different wavelength coverage and sensitivity.}. The non-detections imply that \lya\ emission is
suppressed significantly between $z\sim6$ and $z\sim8$. The 68\%
credible intervals, obtained by integrating the posterior are
$\ep<0.31$ and $\es<0.28$, i.e. \lya\ emission from LBGs is less than
a third than the value at $z\sim6$. The parameters $\ep$ and $\es$ can
be physically interpreted as the average excess optical depth of
Ly-$\alpha$ with respect to $z\sim6$, i.e. $\langle
e^{-\tau_{Ly\alpha}}\rangle$. As expected for a sample of
non-detections, the data are insufficient to distinguish between the
two models. We will thus refer primarily to the patchy model, for
easier comparison with previous work \citep[this is the model
implicitly assumed by][]{Fon++10,Pen++11,Ono++12,Sch++12}.

%= = = = = = = = = = = = = = = = = = = = = = = = = = = = = = = = = = = = = = = = 
\medskip
\medskip

Before discussing the interpretation of our findings we need to
consider the role of contamination. The parameter $\ep$ relates the
number of LBG selected galaxies with \lya\ emission at $z\sim8$ to the
same quantity at $z\sim6$. In order to transform this into a \lya\
optical depth, one has to account for the fraction of contaminants in
both samples

\be
n_{{\rm Ly}\alpha,z=8}=\ep n_{{\rm Ly}\alpha,z=6}\frac{1-f_6}{1-f_8},
\ee
where $f_6$ and $f_8$ are the fraction of contaminants in the $z\sim6$
and $z\sim8$ LBG selected samples, respectively. A simple estimate of
the number of contaminants can be obtained from the posterior
probability distribution functions of the photometric redshifts and by
computing the total probabilities that the source is outside the
fiducial window. This probability is low and does not change our
conclusions in any significant way: \cite{SEO11} estimate $f_6<0.1$
with this method, while for our method it is in the range 0.1-0.2 and
already taken into account by our formalism as described by
\citet{Tre++12}. A more insidious form of contaminants is represented
by the ``unknown unknowns'', like the faint emission line objects
discussed above.

In the case of BoRG this additional contribution is estimated to be
$f_8\sim0.2$, \citep[bringing the total to
0.33-0.42][]{Bradley:2012p23263}.  In the case of the i-dropouts
selected from GOODS \citep{SEO11}, the additional contamination is
probably somewhat less, given the higher quality of the dithering
strategy and larger number of blue bands available.  To be
conservative we thus consider the ratio $(1-f_6)/(1-f_8)$ to be in the
range 1-1.25, that is from equal contamination -- after accounting for
known losses inferred from photo-$z$s -- to higher contamination in
the $z\sim8$ sample.

\medskip
\begin{figure}
\includegraphics[width=0.99\columnwidth]{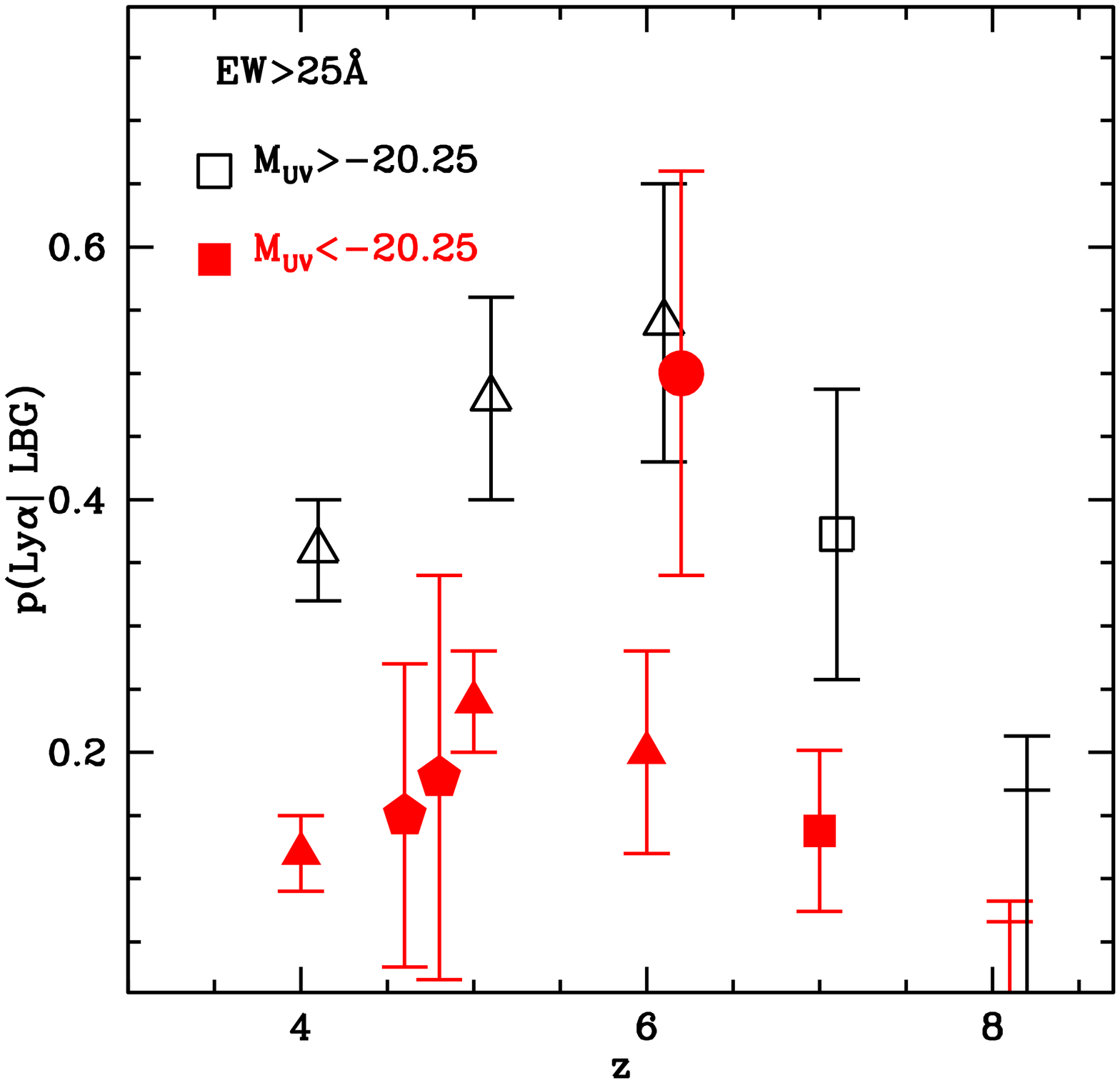}
\caption{Evolution of the fraction of LBGs with Ly-$\alpha>$25 \AA\ equivalent width (rest frame), for bright (filled red symbols) and faint galaxies (open black symbols). Triangles are taken from \citet{SEO11} and \citet{Sch++12}, pentagons from \citet{Mal++12} and the circle is from \citet{Cur++12}. The squares at $z\sim7$ are taken from \citet{Tre++12} and are based on a compilation of data \citep{Fon++10,Van++11,Pen++11,Ono++12,Sch++12}. The upper limits at $z\sim8$ are from this paper. The lower and higher horizontal bars on the upper limits at $z\sim8$ describe the range of uncertainty stemming from contaminants in the photometrically selected LBG sample.
 \label{fig:lyalbg}}
\end{figure} 
\medskip

With this estimate in hand we can proceed to compute the fraction of
LBGs with \lya\ emission above the standard threshold of 25\AA~
equivalent width. Our measurement at $z\sim8$ is shown in
Figure~\ref{fig:lyalbg} together with data from the literature at
lower redshift (see caption). In the patchy model, the fractions for
Y-dropouts are $<0.07-0.08$ for galaxies with $M_{\rm UV}<-20.25$ and
$<0.17-0.21$ for galaxies fainter than this limit (the two numbers are
for minimal and maximal contamination). In the smooth model the same
fractions are $<0.03-0.05$ and $0.06-0.12$. Note that these bounds
include the uncertainty on the $z\sim6$ fraction and thus the
uncertainties on the points beyond $z\sim6$ are correlated. If the
fractions at $z\sim6$ move up/down, so do the points at higher
redshift, but the trend will remain the same. Even considering the
more conservative upper limits from the patchy model, the drop in the
fraction of \lya\ emitters amongst LBG in just 300 Myrs is at least a
factor of $\sim 3$.

There are three possible explanations for our finding, ranging from
the mundane to the very interesting. The first and most mundane
explanation is that samples of Y-dropouts suffer from much higher rate
of contamination than similar LBG samples at lower redshift. A
breakdown of the Lyman Break technique could occur if there were
exotic populations of galaxies which are missing from our current
templates and models used to estimate color-cuts and compute
photo-$z$s. While this cannot be ruled out with present data, it would
certainly be a surprise to see the Lyman Break selection breaking down
so abruptly over a relatively small change in wavelength and
magnitudes. The second explanation could be related to the special
environment of the BoRG galaxies. As expected for the most luminous
galaxies at every redshift, the BoRG sources are bright and strongly
clustered, especially those that we selected for spectroscopic
follow-up \citep{Hil++09,Ove++09}. Thus, we may be comparing galaxies in
proto-clusters with field galaxies, and perhaps this could bias our
interpretation. However, we expect the higher density regions to
completely reionize earlier and therefore to have a smaller
\lya\ optical depth, not larger \citep{B+L04}. Thus this second 
explanation of the large \lya\ optical depth at $z\sim8$ as the result
of a selection bias would also be surprising. The third explanation is
that indeed the average \lya\ optical depth of the universe increases
significantly in this small amount of cosmic time. This third
explanation would be very exciting, implying that we have reached an
epoch where the properties of the intergalactic and circumgalactic
medium (IGM and CGM, respectively) are changing dramatically,
presumably owing to rapid changes in the degree of cosmic reionization
or in the physics of the first generations of star forming regions.

Observationally, the big question is then how do we test these three
hypothesis. Searches for \lya\, to greater depth than ours, would
provide useful information, hopefully including detections.  But also
more and deeper non-detections would certainly help tighten the upper
limits derived here. This is possible with longer MOSFIRE integrations
or by using the WFC3 grism on board the Hubble Space
Telescope. Systematic studies of many gravitationally lensed sources
should be a particularly powerful way to probe very faint sources and
thus also help with testing the second hypothesis
\citep[e.g.,][]{Bra++12}. However, it is probable that some fraction 
of LBGs at $z\sim8$ and above will remain undetected in
\lya\ even with heroic efforts. In order to quantify the amount of contaminants and 
thus test the first and third hypotheses, one needs detections of
other lines or of the continuum.  Hopefully IR lines can be detected
with pointed observations with ALMA \citep{C+W13} although this is
non-trivial \citep{Ouc++13}. Detection of the continuum will be hard
and might require several hours of integrations with the James Webb
Space Telescope \citep{Tre++12} or an extremely large telescope like
the Thirty Meter Telescope, unless the sources are highly magnified by
a foreground gravitational lens.

%========================================================================
\section{Conclusions} \label{sec:conc}

We present MOSFIRE observations of a sample of candidate $z\sim8$
galaxies identified as part of the BoRG Survey. The data are
consistent with noise, setting stringent upper limits on the presence
of emission lines. We carry out a statistical analysis of the
non-detections in the context of our flexible models of purely patchy
and smooth absorption showing that they imply a substantial increase
in \lya\ optical depth $\tau_{Ly\alpha}$ between $z\sim6$ and
$z\sim8$. Quantitatively, our findings can be summarized as follows:

\begin{itemize}
\item At $z\sim8$ the distribution of Ly-$\alpha$ equivalent width 
is significantly reduced with respect to $z\sim6$, by at least a
factor of three (i.e. $\langle e^{-\tau_{Ly\alpha}}\rangle<0.31$ and
$<0.28$ respectively in the patchy and smooth model).
\item The fraction of emitters with equivalent width $>25$\AA\ can 
be computed within our models.  In the patchy model, the fractions for
Y-dropouts are $<0.06-0.08$ for galaxies with $M_{\rm UV}<-20.25$ and
$<0.16-0.21$ for galaxies fainter than this limit (the two numbers are
for minimal and maximal contamination). In the smooth model the same
fractions are $<0.03-0.05$ and $0.06-0.12$.

\end{itemize}

These results extend out to $z\sim8$ in a more dramatic fashion the
increase in \lya\ optical depth seen by previous studies between
$z\sim6$ and $z\sim7$
\citep{Fon++10,Van++11,Pen++11,Sch++12,Ono++12,Tre++12}. This body of
work indicates that the properties of LBG galaxies are evolving over a
very short amount of cosmic time. More spectroscopic data are needed
to characterize this very interesting process further and clarify the
relationship between the vanishing \lya\ emission and cosmic
reionization.

\acknowledgments

Some of the data presented herein were obtained at the W.M. Keck
Observatory.  The authors wish to recognize and acknowledge the very
significant cultural role and reverence that the summit of Mauna Kea
has always had within the indigenous Hawaiian community.  We are
grateful to the MOSFIRE team and the staff at Keck for making these
observations possible, and to N.P.Konidaris for writing the data
reduction pipeline. This paper is also based on observations made with
the NASA/ESA Hubble Space Telescope, obtained at the Space Telescope
Science Institute. We acknowledge support through grants HST-11700,
12572, 12905, and we thank the referee for a constructive report,
which improved the manuscript. T.T. thanks L.Pentericci, E.Vanzella,
M.Giavalisco, and D.Stark for useful conversations.

%======================================================================
\end{document}